\documentclass[10pt, conference, a4paper]{IEEEtran}
\usepackage{amsmath,amsfonts,amssymb,amsbsy, amsthm,ae,aecompl}
\usepackage{algorithm,algpseudocode}
\usepackage[utf8]{inputenc}
\usepackage[english]{babel}
\usepackage{bm}
\usepackage{caption, subcaption}
\usepackage{color}
\usepackage[noadjust]{cite}
\usepackage{epsfig}
\usepackage{enumerate}
\usepackage{float}
\usepackage{fancyhdr}
\usepackage[T1]{fontenc}
\usepackage[acronym,toc,shortcuts]{glossaries}
\usepackage{graphicx}
\usepackage{lastpage}
\usepackage{listings}
\usepackage{lipsum}
\usepackage{dsfont}
\usepackage{multind}
\usepackage[normalem]{ulem}
\usepackage{transparent}
\usepackage{tikz,pgfplots}
\usepackage{times}
\usepackage{verbatim}
\usepackage[all]{xy}

\usepackage{etoolbox}
\AtBeginEnvironment{align}{\setcounter{subeqn}{0}}
\newcounter{subeqn} %

\usetikzlibrary{calc}
\pgfplotsset{
  grid style = {
    dash pattern = on 0.025mm off 0.95mm on 0.025mm off 0mm, 
    line cap = round,
    black,
    line width = 0.5pt
  },
  tick label style={font=\small},
  label style={font=\small},
  legend style={font=\footnotesize},
}

\pgfplotscreateplotcyclelist{laneas4tl}{
cyan!60!black,		solid,\\%
cyan!60!black, 		solid, every mark/.append style={solid,fill=cyan!60!black},mark=o\\%
red!80!black, 		densely dashed, every mark/.append style={solid, fill=lime!80!black},mark=square\\%
red!80!black,		densely dashed,\\%
}

\graphicspath{{figures/}}

\bgroup
\makeglossaries
\newacronym{BS}{BS}{base station}
\newacronym{CDN}{CDN}{content delivery network}
\newacronym{CF}{CF}{collaborative filtering}
\newacronym{CN}{CN}{core network}
\newacronym{CRP}{CRP}{{C}hinese restaurant process}
\newacronym{CS}{CS}{central scheduler}
\newacronym{D2D}{D2D}{device-to-device}
\newacronym{HetNet}{HetNet}{heterogeneous network}
\newacronym{ICIC}{ICIC}{inter-cell interference coordination}
\newacronym{ICN}{ICN}{information-centric network}
\newacronym{LTE}{LTE}{long term evolution}
\newacronym{MIMO}{MIMO}{multiple-input multiple-output}
\newacronym{PPP}{PPP}{{P}oisson point process}
\newacronym{PHY}{PHY}{physical layer}
\newacronym{SBS}{SBS}{small base station}
\newacronym{SINR}{SINR}{signal-to-interference-plus-noise ratio}
\newacronym{SCN}{SCN}{small cell network}
\newacronym{SVD}{SVD}{singular value decomposition}
\newacronym{TL}{TL}{transfer learning}
\newacronym{UT}{UT}{user terminal}
\newacronym{QoS}{QoS}{quality-of-service}
\newacronym{QoE}{QoE}{quality-of-experience}
\newacronym{RAN}{RAN}{radio access network}

\begin{document} 
\title{A Transfer Learning Approach for Cache-Enabled Wireless Networks}
\author{
		\IEEEauthorblockN{Ejder Baştuğ$^{\diamond}$, Mehdi Bennis$^{\star}$ and Mérouane Debbah$^{\diamond,\dagger}$}
		\IEEEauthorblockA{ 
				\vspace{-0.45cm} 
				\\
				\small
				$^{\diamond}$Large Networks and Systems Group (LANEAS), CentraleSupélec, Gif-sur-Yvette, France \\	
				$^{\star}$Centre for Wireless Communications, University of Oulu, Finland \\
				$^{\dagger}$Mathematical and Algorithmic Sciences Lab, Huawei France R\&D, Paris, France \\
				ejder.bastug@centralesupelec.fr, bennis@ee.oulu.fi, merouane.debbah@huawei.com
				\vspace{-0.45cm}
		}
		\thanks{This research has been supported by the ERC Starting Grant 305123 MORE (Advanced Mathematical Tools for Complex Network Engineering), the SHARING project under the Finland grant 128010 and the project BESTCOM.}
}
\IEEEoverridecommandlockouts
\maketitle
\begin{abstract}
Locally caching contents at the network edge constitutes one of the most disruptive approaches in $5$G wireless networks. Reaping the benefits of edge caching hinges on solving a myriad of challenges such as how, what and when to strategically cache contents subject to storage constraints, traffic load, unknown spatio-temporal traffic demands and data sparsity. Motivated by this, we propose a novel \emph{transfer learning}-based caching procedure carried out at each small cell base station. This is done by exploiting the rich contextual information (i.e., users' content viewing history, social ties, etc.) extracted from device-to-device (D2D) interactions, referred to as \emph{source domain}. This prior information is incorporated in the so-called \emph{target domain} where the goal is to  optimally cache strategic contents at the small cells as a function of storage, estimated content popularity, traffic load and backhaul capacity. It is shown that the proposed approach overcomes the notorious data sparsity and cold-start problems, yielding significant gains in terms of users' quality-of-experience (QoE) and backhaul offloading, with gains reaching up to $22\%$ in a setting consisting of four small cell base stations.
\end{abstract}
\vspace{-0.0cm}
\section{Introduction}
Caching at the network edge is one of the five most promising innovations  in $5$G wireless networks \cite{Boccardi2014Five}. Recently, it was shown that caching can significantly offload different segments of the infrastructure including \ac{RAN} and \ac{CN}, by intelligently storing contents closer to the users. As opposed to pushing contents on a \emph{best-effort} basis ignoring end-users' behavior and interactions, we are witnessing an era of truly context-aware and proactive networking  \cite{Bastug2014LivingOnTheEdge}. Undoubtedly, edge caching has taken recent $5$G research activities by storm as evidenced by the recent literature in both academia and industry \cite{Bastug2014LivingOnTheEdge, Golrezaei2012Femtocaching, Bastug2014CacheEnabledExtended, Altieri2014Fundamental, Blasco2014Learning, Poularakis2014Approximation, Hamidouche2014ManyToMany,  MaddahAli2014Fundamental, Paakkonen2013Device, ElBamby2014ContentAware, Liu2014CacheEnabled} (to cite a few). 

Although caching has been well-studied in wired networks, caching over wireless remains in its infancy. The idea of \emph{femtocaching} was proposed in \cite{Golrezaei2012Femtocaching}, in which \glspl{SBS} called \emph{helpers} with low-speed backhaul but high storage units carry out content delivery via short-range transmissions. Randomly distributed \glspl{SBS} with storage capabilities are studied in \cite{Bastug2014CacheEnabledExtended}, characterizing the outage probability and average delivery rate. A stochastic-geometry based caching framework for \ac{D2D} communications is examined in \cite{Altieri2014Fundamental} where  mathematical expressions of local and global fractions of served content requests are given. 
From a game theoretic standpoint, various approaches have been studied such as multi-armed bandits under unknown content popularity \cite{Blasco2014Learning}, many-to-many matching \cite{Hamidouche2014ManyToMany} and joint content-aware user clustering and content caching  \cite{ElBamby2014ContentAware}.
Other works include information-theoretic studies looking at fundamentals of local and global caching gains in \cite{MaddahAli2014Fundamental}, facility location based approximation in \cite{Poularakis2014Approximation}, as well as \ac{MIMO} caching in \cite{Liu2014CacheEnabled}, and coded caching in \cite{Paakkonen2013Device}.

In \cite{Bastug2014LivingOnTheEdge}, by exploiting spatio-social caching coupled with \ac{D2D} communication, we   proposed a novel proactive networking paradigm in which \glspl{SBS} and \glspl{UT} proactively cache contents at the network edge. As a result, the overall performance of the network in terms of users' satisfaction and backhaul offloading was improved. Therein, the \emph{proactive} caching problem assumed non-perfect knowledge of the content popularity matrix, and supervised machine learning  and \ac{CF} techniques were used to estimate the popularity matrix leveraging user-content correlations. Nevertheless, the content popularity matrix remains typically large and sparse with very few users ratings, rendering \ac{CF} learning methods inefficient  mainly due to \emph{data sparseness} and  \emph{cold-start} problems \cite{Lee2012Cf}.

Given the fact that data sparsity and cold-start problems degrade the performance of proactive caching, we leverage the framework of \emph{\ac{TL}} and recent advances in machine learning \cite{Pan2010Transfer}. \ac{TL} is motivated by the fact that in many real-world applications, it is hard or even impossible to collect and label training data to build suitable prediction models. Exploiting available data from other rich information sources such as \ac{D2D} interactions (called as \emph{source domain}),  allows \ac{TL} to  substantially improve the  prediction task in the so-called \emph{target domain}. \ac{TL} has been applied to various data mining problems such as classification and regression \cite{Pan2010Transfer}.
\Ac{TL} methods can be mainly grouped into \emph{inductive}, \emph{transductive} and \emph{unsupervised} \ac{TL} methods depending on the availability of labels in the source and target domains. All these approaches boil down to answering the following fundamental questions: 1) \emph{what} information to transfer? 2) \emph{how} to transfer it? and 3) \emph{when} to transfer it? While "what to transfer" deals with which part of the knowledge should be transferred between domains and tasks, "when to transfer" focuses on the timing of the operations in order to avoid negative transfer, especially when the source and target domains are  uncorrelated. On the other hand, "how to transfer" deals with what kind of information should be transferred between domains and tasks.

The main contribution of this work is to propose a \ac{TL}-based content caching mechanism to maximize the backhaul offloading gains as a function of storage constraints and users' content popularity matrix. This is done by learning and transferring hidden latent features extracted from the source domain to the target domain. In the source domain, we take into account users' \ac{D2D} interactions while accessing/sharing statistics of contents within their social community as prior information in the knowledge transfer. It is shown that the content popularity matrix estimation in the target domain can be significantly  improved instead of \emph{learning from scratch} with unknown users' ratings. To the best of our knowledge, this is perhaps the first contribution of unsupervised transfer learning in cache-enabled small cells.

The rest of the paper is organized as follows. The network model under consideration is provided in Section \ref{sec:networkmodel}, accompanied with the caching problem formulation in both source and target domains. Section \ref{sec:transferlearning} presents the classical \ac{CF}-based caching and that of the proposed  transfer learning. The numerical results capturing the impact of various parameters on the users' satisfaction  and backhaul offloading gains are given in Section \ref{sec:numerical}. We finally conclude and delineate future directions in Section \ref{sec:conclusions}.
%
\vspace{-0.0cm}
\section{Network Model}
\label{sec:networkmodel}
\begin{figure}[ht!]
	\centering
	\includegraphics[width=0.8\columnwidth]{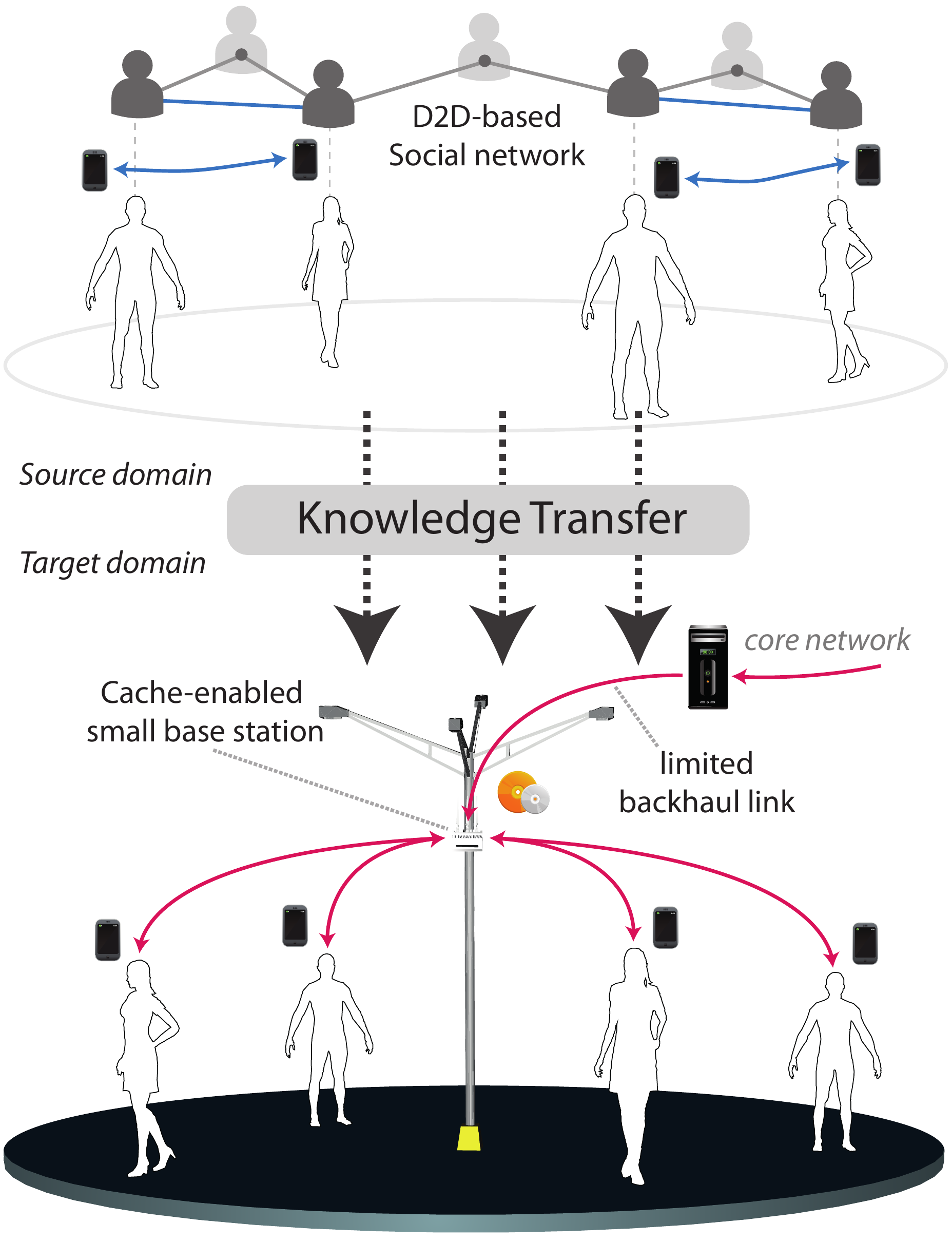}
	\caption{An illustration of the network model which consists of two information systems $S^{(\mathcal{S})}$ and $S^{(\mathcal{T})}$. Due to the lack of prior information in the target domain, the information extracted from users' social interactions and their ratings in the source domain is transferred to the target domain.}
	\label{fig:scenario}
	\vspace{-0.2cm}
\end{figure}

Let us assume an information system denoted by $S^{(\mathcal{S})}$ in the source domain and an information system denoted by $S^{(\mathcal{T})}$ in the target domain. A sketch of the network model is shown in Fig. \ref{fig:scenario}.
\subsection{Target Domain} 
Let us consider a network deployment consisting of $M_{tar}$ \glspl{SBS} from the set $\mathcal{M}_{tar} = \lbrace 1, \ldots, M_{tar} \rbrace$ and $N_{tar}$ \glspl{UT} from the set $\mathcal{N}_{tar}= \lbrace 1, \ldots, N_{tar}\rbrace$. Each \ac{SBS} $m$ is connected to the core network via a limited backhaul link with capacity $0 < C_m < \infty$ and each \ac{SBS} has a total wireless link capacity $C'_m$ for serving its \glspl{UT} in the downlink. We further assume that $\mathbb{E}[C_m] < \mathbb{E}[C'_m]$. \glspl{UT} request contents from a library $\mathcal{F}_{tar} = \lbrace 1, \ldots, F_{tar}\rbrace$, where each content $f$ has a size of $L(f)$ and a bitrate requirement of $B(f)$.
Moreover, we suppose that users' content requests follow a Zipf-like distribution $P_{\mathcal{F}_{tar}}(f), \forall f \in \mathcal{F}_{tar}$ defined as \cite{Breslau1999WebZipf}:
\begin{equation}
	P_{\mathcal{F}_{tar}}(f) = \frac{\Omega}{f^{\alpha}}
\end{equation}
where $ \Omega = \Big(\sum_{i=1}^{F_{tar}}{\frac{1}{i^{\alpha}}}\Big)^{-1} \nonumber$ and $\alpha$ characterizes the steepness of the distribution, reflecting different content popularities. Having such a content popularity in the ordered case, the content popularity matrix for the $m$-th \ac{SBS} at time $t$ is given by $\bold{P}^{m}(t) \in \mathbb{R}^{N_{tar} \times F_{tar}}$ where each entry $P^m_{n,f}(t)$ represent the probability that the $n$-th user requests the $f$-th content. 

In order to avoid any kind of bottleneck during the delivery of users' content requests, we assume that each \ac{SBS} has a finite storage capacity of $S_m$ and caches selected contents from the library $\mathcal{F}_{tar}$. Thus, the amount of requests  \glspl{SBS} satisfy from their local caches is of high importance to avoid peak demands and minimize the latency of content delivery. Our goal is to offload the backhaul while satisfying users' content requests, by pre-fetching strategic contents from the \ac{CN} at suitable times and cache them at the \glspl{SBS}, subject to their storage constraints. To formalize this, suppose that $D$ number of requests from the set $\mathcal{D} = \{1, ..., D\}$ are made by users during $T$ time-slots. Then, a request $d \in \mathcal{D}$ within time window $T$ is served immediately and is said to be \emph{satisfied}, if the rate of delivery is equal or greater than the content bitrate, such that:
\begin{equation}
	\frac{L(f_d)}{\tau'(f_d)  - \tau(f_d)} \ge B(f_d)
\end{equation}
where $f_d$ is the requested content, $L(f_d)$ and $B(f_d)$ are the size and bitrate of the content, $\tau(f_d)$ is the arrival time of the request and $\tau'(f_d)$ the end time delivery. Given these definitions, the users' average \emph{satisfaction ratio} can be expressed as:
\begin{equation}
	\eta(\mathcal{D}) = \frac{1}{D}\sum_{d \in \mathcal{D}}{\mathds{1} \left\lbrace \frac{L(f_d)}{\tau'(f_d)  - \tau(f_d)} \ge B(f_d) \right\rbrace}
\end{equation}
where $\mathds{1}\left\lbrace ... \right\rbrace$ is the indicator function which returns $1$ if the statement holds and $0$ otherwise. Suppose that the instantaneous backhaul rate for the content delivery of request $d$ at time $t$ is given by $R_d(t) \leq C_m$, $\forall m \in \mathcal{M}_{tar}$. Then, the average \emph{backhaul load} is defined as:
\begin{equation}
	\label{eq:bload}
	\rho(\mathcal{D}) = \frac{1}{D}\sum_{d \in \mathcal{D}}{\frac{1}{L(f_d)}\sum_{t=\tau(f_d)}^{\tau'(f_d)}{R_d(t)}}.
\end{equation}
Now, denote ${\bf X}(t) \in \{0, 1\}^{M_{tar}\times F_{tar}}$ as the cache decision matrix of \glspl{SBS}, where $x_{m,f}(t)$ equals $1$ if the $f$-th content is cached at the $m$-th \ac{SBS} at time $t$, and $0$ otherwise. Therefore, the backhaul offloading problem can be formally expressed as:
\begin{align}
\label{eq:problem1}
&	\underset{{\bf X}(t), {\bf P}^m(t)}{\text{minimize}}	&	& \rho(\mathcal{D}) 			     \\
&	\text{subject to}	& & L_{\text{min}} \leq L(f_d) \leq L_{\text{max}},	\hspace{2.2cm}   \forall d \in \mathcal{D}, \nonumber \\
&						& & B_{\text{min}} \leq B(f_d) \leq B_{\text{max}}, 	\hspace{2.15cm}   \forall d \in \mathcal{D},\nonumber \\
&						& & R_d(t) \leq C_m, \hspace{1.2cm} \forall t, \forall d \in \mathcal{D}, \forall m \in \mathcal{M}_{tar},\nonumber \\
&						& & R'_d(t) \leq C'_m, \hspace{1.2cm} \forall t, \forall d \in \mathcal{D}, \forall m \in \mathcal{M}_{tar}, \nonumber \\
&						& & \sum_{f \in \mathcal{F}_{tar}} L(f)x_{m,f}(t) \leq S_m, 	\hspace{0.4cm}  \forall t,  \forall m \in \mathcal{M}_{tar},\nonumber \\
&						& & \sum_{n \in \mathcal{N}_{tar}}\sum_{f \in \mathcal{F}_{tar}}{P^m_{n,f}(t)} = 1, 		   \hspace{0.45cm} \forall t, \forall m \in \mathcal{M}_{tar}, \nonumber \\
&						& & x_{m,f}(t) \in \{0, 1\},  \hspace{0.2cm} 		 \forall t, \forall f \in \mathcal{F}_{tar}, \forall m \in \mathcal{M}_{tar}, \nonumber \\
&						& & \eta_{\text{min}} \leq \eta(\mathcal{D}) 		   \nonumber
\end{align}
where $R'_d(t)$ is the instantaneous wireless link rate for request $d$ and $\eta_{\text{min}}$ is the minimum target satisfaction ratio respectively. In order to solve this problem, a joint optimization of the cache decision ${\bf X}(t)$ and the content popularity matrix estimation ${\bf P}^m(t)$ is needed. Moreover, solving (\ref{eq:problem1}) is very  challenging due to: 
\begin{itemize}
\item[i)] limited backhaul and wireless link capacity as well as the limited storage capacity of \glspl{SBS}, 
\item[ii)] large number of users with unknown ratings and library size, 
\item[iii)] \glspl{SBS} need to track, learn and estimate users' content popularity/rating matrix $\bold{P}^m(t)$ for cache decision while dealing with data sparsity.
\end{itemize}
For simplicity, we drop now the index of the \glspl{SBS} and assume that the content popularity is stationary during $T$ time slots, thus $\bold{P}^m(t)$ is denoted as $\bold{P}_{tar}$. Moreover, for sake of exposition, we restrict ourselves to caching policies in which the contents are stored during the peak-off hours, thus ${\bf X}(t)$ remains fixed during the content delivery and represented as ${\bf X}$. In the following, we examine the source domain which we exploit when dealing with the sparsity of $\bold{P}_{tar}$ in the target domain.
\subsection{Source Domain} 
As advocated in \cite{Bastug2014LivingOnTheEdge}, we leverage the existence of a \ac{D2D}-based social network overlay made of users' interactions within their social communities, referred as the \emph{source domain} in the sequel. Specifically, this source domain contains the behaviour of users' interactions within their social communities, modelled as a \ac{CRP} \cite{Griffiths2011CRP}. This constitutes the prior information used in the transfer learning procedure.

In the \ac{CRP} with parameter $\beta$, every customer selects an occupied table with a probability proportional to the number of occupants, and selects the next vacant table with probability proportional to $\beta$. More precisely, the first customer selects the first table with probability $\frac{\beta}{\beta}=1$. The second customer selects the first table with probability $\frac{1}{1+\beta}$, and the second table with probability $\frac{\beta}{1+\beta}$. After the second customer selects the second table, the third customer chooses the first table with probability $\frac{1}{2+\beta}$, the second table with probability $\frac{1}{2+\beta}$ and the third table with probability $\frac{\beta}{2+\beta}$. This stochastic Dirichlet process continues until all customers select their seats, defining a distribution over allocation of customers to tables.

In this regard, the content dissemination in the social network is analogous to the table selection in a \ac{CRP}. If we view this network as a \ac{CRP}, the contents as the large number of tables, and users as the customers, we can make an analogy between the content dissemination and the \ac{CRP}. First, suppose that there exist $N_{D2D}$ users in this network. Let $F_{D2D} = F_0+F_h$ be the total number of contents in which $F_h$ represents the number of contents with viewing histories and $F_0$ is the number of contents without history. Denote also $\bold{Z}_{D2D} \in \{0, 1\}^{N_{D2D} \times F_{D2D}}$ as a random binary matrix indicating which contents are selected by each user, where $z_{n,f}=1$ if the $n$-th user selects the $f$-th content and $0$ otherwise. Then, it can be shown that \cite{Griffiths2011CRP}:
\begin{equation}
P(\bold{Z}_{D2D})=\frac{\beta^{F_h}\Gamma(\beta)}{\Gamma(\beta+N_{D2D})}\prod_{f=1}^{F_h} (m_{f}-1)!
\end{equation}
where $\Gamma(.)$ is the Gamma function, $m_f$ is the number of users assigned to content $f$ (i.e., viewing history) and $F_h$ is the number of contents with viewing histories with $m_f>0$.

In the target domain, the caching problem boils down to estimating the content popularity matrix which is assumed to be largely unknown, yielding degraded performance (i.e., very low cache hit ratios, slow convergence, etc.). Moreover, this degradation can be more severe in cases where the number of users and library size is extremely large. Therefore, in order to handle these issues and cache contents more efficiently, we propose a novel proactive caching procedure using transfer learning which exploits the rich contextual information extracted from users' social interactions. This caching procedure is shown to yield more backhaul  offloading gains compared to a number of baselines, including random caching and the classical \ac{CF}-based estimation methods \cite{Bastug2014LivingOnTheEdge}. 

\vspace{-0.0cm}
\section{Transfer Learning: Boosting Content Popularity Matrix Estimation}
\label{sec:transferlearning}
First, we start by explaining the classical \ac{CF}-based learning, then detail our proposed \ac{TL} solution.
\subsection{Classical CF-based Learning}
The classical \ac{CF}-based estimation procedure is composed of a training and prediction phase. In the training part, the goal is to estimate the content popularity matrix $\bold{P}_{tar} \in \mathbb{R}^{N_{tar} \times F_{tar}}$, where each \ac{SBS} constructs a model based on the already available information (i.e., users' content ratings). Let $\mathcal{N}_{tar}$ and $\mathcal{F}_{tar}$ represent the set of users and contents associated with $N_{tar}$ users and $F_{tar}$ contents. In particular, $\bold{P}_{tar}$  with entries $P_{tar,ij}$ is the  (sparse) content popularity matrix in the target domain. $\mathcal{R}_{tar}=\lbrace (i,j,r): r = P_{tar,ij},  P_{tar,ij} \neq 0 \rbrace$ denotes the set of known user ratings. In the prediction phase, in order to predict the unobserved ratings in $\mathcal{N}_{tar}$, low-rank matrix factorization techniques are used to estimate the unknown entries of $\bold{P}_{tar}$. The objective here is to construct a $k$-rank approximate  popularity matrix $\bold{P}_{tar} \approx  \bold{N}_{tar}^T\bold{F}_{tar}$, where the factor matrices $\bold{N}_{tar} \in \mathbb{R}^{k \times N_{tar}}$ and $\bold{F}_{tar} \in \mathbb{R}^{k \times F{tar}}$ are learned by minimizing the following cost function:
\begin{align}
	\label{eq:classical}
	& & \underset{(i,j) \in \bold{P}_{tar}}{\text{minimize}}	& & \sum_{(i,j) \in \bold{P}_{tar}} \Big(\bold{n}_i^T\bold{f}_j-P_{tar,ij}\Big)^2 + \\ \nonumber 	
	& & & & \mu\Big(||\bold{N}_{tar}||^2_F+||\bold{F}_{tar}||^2_F\Big) 		
\end{align}
where the sum is over the ($i$,$j$) user/content pairs in the training set. In addition, $\bold{n}_i$ and $\bold{f}_j$ represent the $i$-th and $j$-th columns of $\bold{N}_{tar}$ and $\bold{F}_{tar}$ respectively, and $||.||^2_F$ denotes the Frobenius norm. In $\eqref{eq:classical}$, the parameter $\mu$ provides a balance between regularization and fitting training data. Unfortunately, users may rate very few contents, causing $\bold{P}_{tar}$ to be extremely sparse, and thus   $\eqref{eq:classical}$ suffers from severe over-fitting issues and engenders poor performance.
\subsection{TL-based Content Caching}
To alleviate data sparsity, solving $\eqref{eq:classical}$ can be done more efficiently by exploiting and transferring the vast amount of available user-content ratings (i.e., prior information) from a different-yet-related source domain. Formally speaking, let us denote the source domain as $S^{(\mathcal{S})}$, and assume that this domain is associated with a set of $N_{D2D}$ users and $F_{D2D}$ contents denoted by $\mathcal{N}_{D2D}$ and $\mathcal{F}_{D2D}$ respectively. Additionally, the user-content popularity matrix in the source domain is given by matrix $\bold{P}_{D2D} \in \mathbb{R}^{N_{D2D} \times F_{D2D}}$ and likewise let $\mathcal{R}_{D2D}=\lbrace (i,j,r): r = P_{D2D,ij},  P_{D2D,ij} \neq 0\rbrace$ represent the set of observed user ratings in the source domain. The underlying principle of the proposed approach is to smartly "borrow" carefully-chosen user social  behavior information from $S^{(\mathcal{S})}$ to better learn $S^{(\mathcal{T})}$.

The transfer learning procedure from $S^{(\mathcal{S})}$ to  $S^{(\mathcal{T})}$ is composed of two interrelated phases. In the first phase, a content \emph{correspondence} is  established in order to identify similarly-rated contents in both source and target domains. In the second phase, an optimization problem is formulated by combining the source and target domains for \emph{knowledge transfer}, to jointly learn the popularity matrix $\bold{P}_{tar}$ in the target domain. In this regard, we suppose that both source and target domains correspond to one information system $s \in \lbrace S^{(\mathcal{S})}, S^{(\mathcal{T})} \rbrace$, that is made of $N_s$ users and $F_s$ contents given by $\mathcal{N}_s$ and $\mathcal{F}_s$ respectively. In each system $s$, we observe $\bold{P}_{s}$ with entries $P_{s,ij}$. Let $\mathcal{R}_{s}=\lbrace (i,j,r): r=P_{s,ij},  P_{s,ij} \neq 0\rbrace$ represent the set of observed user ratings in each system and the set of \emph{shared contents} is given by $\mathcal{\tilde{F}}$. Moreover, let $\mathcal{N}^*=\mathcal{N}_{D2D} \cup \mathcal{N}_{tar}$ and $\mathcal{F}^*=\mathcal{F}_{D2D} \cup \mathcal{F}_{tar}$ be the union of the collections of users and contents, respectively, where $N^*=|\mathcal{N}^*|$ and $F^*=|\mathcal{F}^*|$  represent the total number of unique users and contents in the union of both systems.

\begin{figure}[ht!]
	\centering
	\resizebox{.9\linewidth}{!}{
		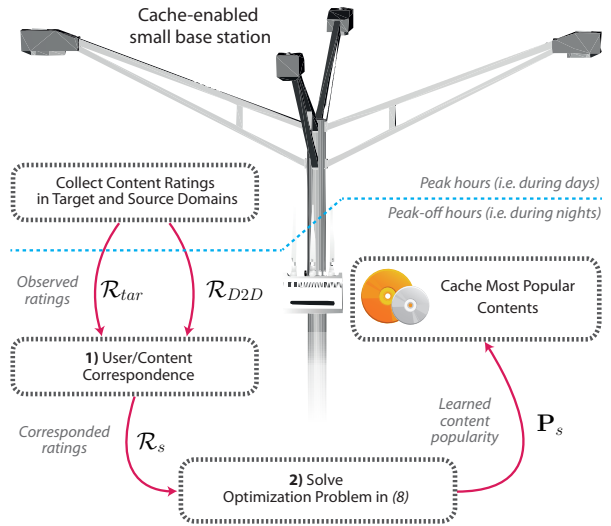
	}
	\caption{An illustration of the proposed \ac{TL}-based caching procedure.}
	\label{fig:procedure}
	\vspace{-0.2cm}
\end{figure}

In the proposed \ac{TL} approach, we model the users $\mathcal{N}^*$  and contents $\mathcal{F}^*$ by a user factor matrix $\bold{N} \in \mathbb{R}^{k \times N^*}$ and  a content factor matrix $\bold{F} \in \mathbb{R}^{k \times F^*}$, where the $i$-th and $j$-th columns of these matrices are given by $\bold{n}_i$ and $\bold{f}_j$, respectively. The aim is to approximate the  popularity matrix $\bold{P}_{s} \approx \bold{N}_{s}^T\bold{F}_{s}$ by jointly learning the factor matrices $\bold{N}$ and $\bold{F}$. This is formally done by minimizing the following cost function:
\begin{align}
	\label{eq:transferlearning}
	& & \underset{(i,j) \in \bold{P}_{s}}{\text{minimize}}	& & \sum_{s}\Big(\alpha_s \sum_{(i,j) \in \bold{P}_{s}} \Big(\bold{n}_i^T\bold{f}_j - P_{s,ij}\Big)^2\Big)+ \\ \nonumber 
	& & & & \mu\Big(||\bold{N}||^2_F+||\bold{F}||^2_F\Big) 		
\end{align}
where the parameter $\alpha_s$ is the weight of each system. By doing so, $\bold{P}_{D2D}$ and $\bold{P}_{tar}$ are jointly factorized, and thus the set of factor matrices $\bold{F}_{D2D}$ and $\bold{F}_{tar}$ become interdependent as the features of a shared content are similar for knowledge sharing. A practical \ac{TL}-based caching procedure is sketched in Fig. \ref{fig:procedure}.
\vspace{-0.0cm}
\section{Numerical Results and Discussion}
\label{sec:numerical}
\begin{figure*}[ht!]
\centering
\begin{subfigure}[t]{.24\textwidth}
\begin{tikzpicture}
	\begin{axis}[
		width=\textwidth,
		grid = major,
		cycle list name=laneas4tl,
	  legend columns=4,
		legend entries={Ground Truth, Random, Collaborative Filtering, Transfer Learning},
		legend cell align=center,
	  	legend style ={font=\normalsize},
		legend to name=namedmethods4tl,
		xlabel=Storage ratio,
		ylabel=Satisfaction ratio]
		
	\addplot+[only marks, mark size=1, semithick] plot coordinates {
		(1.6306e-16,	0)
		(0.083333,	0.055775)
		(0.16667,	0.15055)
		(0.25,	0.25476)
		(0.33333,	0.36852)
		(0.41667,	0.49761)
		(0.5,	0.6166)
		(0.58333,	0.70953)
		(0.66667,	0.77002)
		(0.75,	0.81754)
		(0.83333,	0.84953)
		(0.91667,	0.86584)
		(1,	0.86727)
	}; 	

	\addplot+[only marks, mark size=1, semithick] plot coordinates {
		(1.6306e-16,	0.0039591)
		(0.083333,	0.0081005)
		(0.16667,	0.018463)
		(0.25,	0.035457)
		(0.33333,	0.059173)
		(0.41667,	0.087242)
		(0.5,	0.12394)
		(0.58333,	0.1807)
		(0.66667,	0.26389)
		(0.75,	0.37655)
		(0.83333,	0.51684)
		(0.91667,	0.68467)
		(1,	0.88037)
	}; 	

	\addplot+[only marks, mark size=1, semithick] plot coordinates {
		(1.6306e-16,	0.01562)
		(0.083333,	0.00098931)
		(0.16667,	0.002805)
		(0.25,	0.022282)
		(0.33333,	0.059518)
		(0.41667,	0.10901)
		(0.5,	0.19599)
		(0.58333,	0.33639)
		(0.66667,	0.49964)
		(0.75,	0.63124)
		(0.83333,	0.7466)
		(0.91667,	0.84514)
		(1,	0.92303)
	}; 	

	\addplot+[only marks, mark size=1, semithick] plot coordinates {
		(1.6306e-16,	0.0071058)
		(0.083333,	0.02803)
		(0.16667,	0.066809)
		(0.25,	0.12391)
		(0.33333,	0.19905)
		(0.41667,	0.29148)
		(0.5,	0.41551)
		(0.58333,	0.55201)
		(0.66667,	0.66478)
		(0.75,	0.75133)
		(0.83333,	0.8168)
		(0.91667,	0.86041)
		(1,	0.88141)
	}; 	

	\end{axis}
\end{tikzpicture}
\end{subfigure}
%
\begin{subfigure}[t]{.24\textwidth}
\begin{tikzpicture}
	\begin{axis}[
		width=\textwidth,
		grid = major,
		scaled y ticks={base 10:1},
		cycle list name=laneas4tl,
		legend cell align=left,
		legend style ={legend pos=south east},
		xlabel=CRP concentration parameter,
		ylabel=Satisfaction ratio]
		
	\addplot+[only marks, mark size=1, semithick] plot coordinates {
		(0.02,	0.14479)
		(0.10167,	0.11554)
		(0.18333,	0.093839)
		(0.265,	0.079207)
		(0.34667,	0.072315)
		(0.42833,	0.073788)
		(0.51,	0.072204)
		(0.59167,	0.070654)
		(0.67333,	0.069523)
		(0.755,	0.068622)
		(0.83667,	0.067947)
		(0.91833,	0.067514)
		(1,	0.067316)
	}; 	

	\addplot+[only marks, mark size=1, semithick] plot coordinates {
		(0.02,	0.029142)
		(0.10167,	0.028691)
		(0.18333,	0.028267)
		(0.265,	0.027876)
		(0.34667,	0.027507)
		(0.42833,	0.027159)
		(0.51,	0.02705)
		(0.59167,	0.026929)
		(0.67333,	0.026746)
		(0.755,	0.026656)
		(0.83667,	0.02662)
		(0.91833,	0.02663)
		(1,	0.026696)
	}; 	

	\addplot+[only marks, mark size=1, semithick] plot coordinates {
		(0.02,	0.029426)
		(0.10167,	0.029276)
		(0.18333,	0.029078)
		(0.265,	0.028848)
		(0.34667,	0.028598)
		(0.42833,	0.028214)
		(0.51,	0.027907)
		(0.59167,	0.027886)
		(0.67333,	0.027872)
		(0.755,	0.027929)
		(0.83667,	0.028087)
		(0.91833,	0.028327)
		(1,	0.028644)
	}; 	

	\addplot+[only marks, mark size=1, semithick] plot coordinates {
		(0.02,	0.088789)
		(0.10167,	0.070172)
		(0.18333,	0.056386)
		(0.265,	0.047127)
		(0.34667,	0.042789)
		(0.42833,	0.043757)
		(0.51,	0.043144)
		(0.59167,	0.042807)
		(0.67333,	0.042304)
		(0.755,	0.041758)
		(0.83667,	0.041363)
		(0.91833,	0.041077)
		(1,	0.040877)
	}; 	

	\end{axis}
\end{tikzpicture}
\end{subfigure}
%
\begin{subfigure}[t]{.24\textwidth}
\begin{tikzpicture}
	\begin{axis}[
		width=\textwidth,
		grid = major,
		scaled y ticks={base 10:1},
		cycle list name=laneas4tl,
		legend cell align=left,
		legend style ={legend pos=south east},
		xlabel=Traffic intensity,
		ylabel=Satisfaction ratio]
		
	\addplot+[only marks, mark size=1, semithick] plot coordinates {
		(0.33333,	0.14303)
		(0.38889,	0.12159)
		(0.44444,	0.10362)
		(0.5,	0.088967)
		(0.55556,	0.077658)
		(0.61111,	0.070249)
		(0.66667,	0.064839)
		(0.72222,	0.05983)
		(0.77778,	0.055109)
		(0.83333,	0.051023)
		(0.88889,	0.047514)
		(0.94444,	0.04455)
		(1,	0.042137)
	}; 	

	\addplot+[only marks, mark size=1, semithick] plot coordinates {
		(0.33333,	0.055403)
		(0.38889,	0.047328)
		(0.44444,	0.040503)
		(0.5,	0.034873)
		(0.55556,	0.030474)
		(0.61111,	0.02749)
		(0.66667,	0.024957)
		(0.72222,	0.023075)
		(0.77778,	0.021933)
		(0.83333,	0.020779)
		(0.88889,	0.019759)
		(0.94444,	0.018894)
		(1,	0.018143)
	}; 	

	\addplot+[only marks, mark size=1, semithick] plot coordinates {
		(0.33333,	0.060383)
		(0.38889,	0.054125)
		(0.44444,	0.048358)
		(0.5,	0.043133)
		(0.55556,	0.038541)
		(0.61111,	0.034153)
		(0.66667,	0.030305)
		(0.72222,	0.02768)
		(0.77778,	0.025309)
		(0.83333,	0.023323)
		(0.88889,	0.021788)
		(0.94444,	0.020675)
		(1,	0.01998)
	}; 	

	\addplot+[only marks, mark size=1, semithick] plot coordinates {
		(0.33333,	0.088075)
		(0.38889,	0.07443)
		(0.44444,	0.063025)
		(0.5,	0.053789)
		(0.55556,	0.046824)
		(0.61111,	0.042305)
		(0.66667,	0.038717)
		(0.72222,	0.03582)
		(0.77778,	0.033005)
		(0.83333,	0.03054)
		(0.88889,	0.028435)
		(0.94444,	0.026688)
		(1,	0.025313)
	}; 	

	\end{axis}
\end{tikzpicture}
\end{subfigure}
%
\begin{subfigure}[t]{.24\textwidth}
\begin{tikzpicture}
	\begin{axis}[
		width=\textwidth,
		grid = major,
		cycle list name=laneas4tl,
		legend cell align=left,
		legend style ={legend pos=south east},
		xlabel=Backhaul capacity,
		ylabel=Satisfaction ratio]
		
	\addplot+[only marks, mark size=1, semithick] plot coordinates {
		(0.125,		0.040758)
		(0.19792,	0.040858)
		(0.27083,	0.040898)
		(0.34375,	0.043401)
		(0.41667,	0.048362)
		(0.48958,	0.055099)
		(0.5625,	0.065306)
		(0.63542,	0.082714)
		(0.70833,	0.10923)
		(0.78125,	0.14175)
		(0.85417,	0.18079)
		(0.92708,	0.22639)
		(1,	0.27841)
	}; 	

	\addplot+[only marks, mark size=1, semithick] plot coordinates {
		(0.125,	0.018257)
		(0.19792,	0.018101)
		(0.27083,	0.018254)
		(0.34375,	0.018725)
		(0.41667,	0.019512)
		(0.48958,	0.020577)
		(0.5625,	0.021978)
		(0.63542,	0.023918)
		(0.70833,	0.026378)
		(0.78125,	0.030628)
		(0.85417,	0.036537)
		(0.92708,	0.044043)
		(1,	0.053209)
	}; 	

	\addplot+[only marks, mark size=1, semithick] plot coordinates {
		(0.125,	0.019021)
		(0.19792,	0.018914)
		(0.27083,	0.019287)
		(0.34375,	0.020146)
		(0.41667,	0.021476)
		(0.48958,	0.023274)
		(0.5625,	0.025536)
		(0.63542,	0.028828)
		(0.70833,	0.033597)
		(0.78125,	0.042611)
		(0.85417,	0.055548)
		(0.92708,	0.072269)
		(1,	0.092919)
	}; 	

	\addplot+[only marks, mark size=1, semithick] plot coordinates {
		(0.125,	0.025997)
		(0.19792,	0.026149)
		(0.27083,	0.02695)
		(0.34375,	0.028446)
		(0.41667,	0.030618)
		(0.48958,	0.033233)
		(0.5625,	0.036481)
		(0.63542,	0.04212)
		(0.70833,	0.053078)
		(0.78125,	0.070347)
		(0.85417,	0.09334)
		(0.92708,	0.12211)
		(1,	0.15676)
	}; 	

	\end{axis}
\end{tikzpicture}
\end{subfigure}
\\
\vspace{0.4cm}
\begin{subfigure}[t]{.24\textwidth}
\begin{tikzpicture}
	\begin{axis}[
		width=\textwidth,
		grid = major,
		cycle list name=laneas4tl,
		legend cell align=left,
		legend style ={legend pos=south east},
		xlabel=Storage ratio,
		ylabel=Backhaul load]
		
	\addplot+[only marks, mark size=1, semithick] plot coordinates {
		(0.33333,	1-0.3638)
		(0.38889,	1-0.36104)
		(0.44444,	1-0.35927)
		(0.5,	1-0.35842)
		(0.55556,	1-0.3585)
		(0.61111,	1-0.35986)
		(0.66667,	1-0.36148)
		(0.72222,	1-0.36225)
		(0.77778,	1-0.36252)
		(0.83333,	1-0.36298)
		(0.88889,	1-0.3634)
		(0.94444,	1-0.36381)
		(1,	1-0.36428)
	}; 	

	\addplot+[only marks, mark size=1, semithick] plot coordinates {
		(0.33333,	1-0.18436)
		(0.38889,	1-0.18331)
		(0.44444,	1-0.18261)
		(0.5,	1-0.18223)
		(0.55556,	1-0.18221)
		(0.61111,	1-0.1826)
		(0.66667,	1-0.18279)
		(0.72222,	1-0.18303)
		(0.77778,	1-0.18348)
		(0.83333,	1-0.18374)
		(0.88889,	1-0.18387)
		(0.94444,	1-0.18387)
		(1,	1-0.18374)
	}; 	

	\addplot+[only marks, mark size=1, semithick] plot coordinates {
		(0.33333,	1-0.20408)
		(0.38889,	1-0.20553)
		(0.44444,	1-0.20614)
		(0.5,	1-0.20602)
		(0.55556,	1-0.20518)
		(0.61111,	1-0.20295)
		(0.66667,	1-0.20178)
		(0.72222,	1-0.20174)
		(0.77778,	1-0.20209)
		(0.83333,	1-0.20288)
		(0.88889,	1-0.2041)
		(0.94444,	1-0.20571)
		(1,	1-0.20772)
	}; 	

	\addplot+[only marks, mark size=1, semithick] plot coordinates {
		(0.33333,	1-0.25599)
		(0.38889,	1-0.25305)
		(0.44444,	1-0.25069)
		(0.5,	1-0.24896)
		(0.55556,	1-0.24783)
		(0.61111,	1-0.24723)
		(0.66667,	1-0.24821)
		(0.72222,	1-0.24889)
		(0.77778,	1-0.24837)
		(0.83333,	1-0.24873)
		(0.88889,	1-0.24944)
		(0.94444,	1-0.25047)
		(1,	1-0.25194)
	}; 	

	\end{axis}
\end{tikzpicture}
\end{subfigure}
%
\begin{subfigure}[t]{.24\textwidth}
\begin{tikzpicture}
	\begin{axis}[
		width=\textwidth,
		grid = major,
		cycle list name=laneas4tl,
		legend cell align=left,
		legend style ={legend pos=south east},
		xlabel=CRP concentration parameter,
		ylabel=Backhaul load]
		
	\addplot+[only marks, mark size=1, semithick] plot coordinates {
		(0.02,	1-0.54433)
		(0.10167,	1-0.48106)
		(0.18333,	1-0.4332)
		(0.265,	1-0.39978)
		(0.34667,	1-0.382)
		(0.42833,	1-0.38136)
		(0.51,	1-0.37656)
		(0.59167,	1-0.37234)
		(0.67333,	1-0.36913)
		(0.755,	1-0.3664)
		(0.83667,	1-0.36424)
		(0.91833,	1-0.3627)
		(1,	1-0.36175)
	}; 	

	\addplot+[only marks, mark size=1, semithick] plot coordinates {
		(0.02,	1-0.1878)
		(0.10167,	1-0.18713)
		(0.18333,	1-0.18634)
		(0.265,	1-0.18546)
		(0.34667,	1-0.18447)
		(0.42833,	1-0.18332)
		(0.51,	1-0.1829)
		(0.59167,	1-0.18264)
		(0.67333,	1-0.18216)
		(0.755,	1-0.18182)
		(0.83667,	1-0.18161)
		(0.91833,	1-0.18153)
		(1,	1-0.18159)
	}; 	

	\addplot+[only marks, mark size=1, semithick] plot coordinates {
		(0.02,	1-0.19004)
		(0.10167,	1-0.19426)
		(0.18333,	1-0.19767)
		(0.265,	1-0.20022)
		(0.34667,	1-0.20178)
		(0.42833,	1-0.20269)
		(0.51,	1-0.20259)
		(0.59167,	1-0.20153)
		(0.67333,	1-0.20108)
		(0.755,	1-0.20048)
		(0.83667,	1-0.19969)
		(0.91833,	1-0.19879)
		(1,	1-0.19773)
	}; 	

	\addplot+[only marks, mark size=1, semithick] plot coordinates {
		(0.02,	1-0.41402)
		(0.10167,	1-0.35745)
		(0.18333,	1-0.31469)
		(0.265,	1-0.28498)
		(0.34667,	1-0.26932)
		(0.42833,	1-0.26859)
		(0.51,	1-0.26577)
		(0.59167,	1-0.26454)
		(0.67333,	1-0.26298)
		(0.755,	1-0.26174)
		(0.83667,	1-0.2611)
		(0.91833,	1-0.26096)
		(1,	1-0.26129)
	}; 	

	\end{axis}
\end{tikzpicture}
\end{subfigure}
%
\begin{subfigure}[t]{.24\textwidth}
\begin{tikzpicture}
	\begin{axis}[
		width=\textwidth,
		grid = major,
		cycle list name=laneas4tl,
		legend cell align=left,
		legend style ={legend pos=south east},
		xlabel=Traffic intensity,
		ylabel=Backhaul load]
		
	\addplot+[only marks, mark size=1, semithick] plot coordinates {
		(0.33333,	1-0.3638)
		(0.38889,	1-0.36104)
		(0.44444,	1-0.35927)
		(0.5,	1-0.35842)
		(0.55556,	1-0.3585)
		(0.61111,	1-0.35986)
		(0.66667,	1-0.36148)
		(0.72222,	1-0.36225)
		(0.77778,	1-0.36252)
		(0.83333,	1-0.36298)
		(0.88889,	1-0.3634)
		(0.94444,	1-0.36381)
		(1,	1-0.36428)
	}; 	

	\addplot+[only marks, mark size=1, semithick] plot coordinates {
		(0.33333,	1-0.18436)
		(0.38889,	1-0.18331)
		(0.44444,	1-0.18261)
		(0.5,	1-0.18223)
		(0.55556,	1-0.18221)
		(0.61111,	1-0.1826)
		(0.66667,	1-0.18279)
		(0.72222,	1-0.18303)
		(0.77778,	1-0.18348)
		(0.83333,	1-0.18374)
		(0.88889,	1-0.18387)
		(0.94444,	1-0.18387)
		(1,	1-0.18374)
	}; 	

	\addplot+[only marks, mark size=1, semithick] plot coordinates {
		(0.33333,	1-0.20408)
		(0.38889,	1-0.20553)
		(0.44444,	1-0.20614)
		(0.5,	1-0.20602)
		(0.55556,	1-0.20518)
		(0.61111,	1-0.20295)
		(0.66667,	1-0.20178)
		(0.72222,	1-0.20174)
		(0.77778,	1-0.20209)
		(0.83333,	1-0.20288)
		(0.88889,	1-0.2041)
		(0.94444,	1-0.20571)
		(1,	1-0.20772)
	}; 	

	\addplot+[only marks, mark size=1, semithick] plot coordinates {
		(0.33333,	1-0.25599)
		(0.38889,	1-0.25305)
		(0.44444,	1-0.25069)
		(0.5,	1-0.24896)
		(0.55556,	1-0.24783)
		(0.61111,	1-0.24723)
		(0.66667,	1-0.24821)
		(0.72222,	1-0.24889)
		(0.77778,	1-0.24837)
		(0.83333,	1-0.24873)
		(0.88889,	1-0.24944)
		(0.94444,	1-0.25047)
		(1,	1-0.25194)
	}; 	

	\end{axis}
\end{tikzpicture}
\end{subfigure}
%
\begin{subfigure}[t]{.24\textwidth}
\begin{tikzpicture}
	\begin{axis}[
		width=\textwidth,
		grid = major,
		cycle list name=laneas4tl,
		legend cell align=left,
		legend style ={legend pos=south east},
		xlabel=Backhaul capacity,
		ylabel=Backhaul load]
		
	\addplot+[only marks, mark size=1, semithick] plot coordinates {
		(0.125,	1-0.35503)
		(0.19792,	1-0.35503)
		(0.27083,	1-0.35503)
		(0.34375,	1-0.35503)
		(0.41667,	1-0.35503)
		(0.48958,	1-0.35503)
		(0.5625,	1-0.35503)
		(0.63542,	1-0.35503)
		(0.70833,	1-0.35503)
		(0.78125,	1-0.35503)
		(0.85417,	1-0.35503)
		(0.92708,	1-0.35503)
		(1,	1-0.35503)
	}; 	

	\addplot+[only marks, mark size=1, semithick] plot coordinates {
		(0.125,	1-0.18774)
		(0.19792,	1-0.18774)
		(0.27083,	1-0.18774)
		(0.34375,	1-0.18774)
		(0.41667,	1-0.18774)
		(0.48958,	1-0.18774)
		(0.5625,	1-0.18774)
		(0.63542,	1-0.18774)
		(0.70833,	1-0.18774)
		(0.78125,	1-0.18774)
		(0.85417,	1-0.18774)
		(0.92708,	1-0.18774)
		(1,	1-0.18774)
	}; 	

	\addplot+[only marks, mark size=1, semithick] plot coordinates {
		(0.125,	1-0.20051)
		(0.19792,	1-0.20112)
		(0.27083,	1-0.20151)
		(0.34375,	1-0.20168)
		(0.41667,	1-0.20166)
		(0.48958,	1-0.20147)
		(0.5625,	1-0.20136)
		(0.63542,	1-0.20177)
		(0.70833,	1-0.20188)
		(0.78125,	1-0.20156)
		(0.85417,	1-0.20119)
		(0.92708,	1-0.20063)
		(1,	1-0.19985)
	}; 	

	\addplot+[only marks, mark size=1, semithick] plot coordinates {
		(0.125,	1-0.25135)
		(0.19792,	1-0.25215)
		(0.27083,	1-0.25255)
		(0.34375,	1-0.25255)
		(0.41667,	1-0.25216)
		(0.48958,	1-0.25139)
		(0.5625,	1-0.25007)
		(0.63542,	1-0.24943)
		(0.70833,	1-0.25013)
		(0.78125,	1-0.25075)
		(0.85417,	1-0.25151)
		(0.92708,	1-0.25243)
		(1,	1-0.25342)
	}; 	

	\end{axis}
\end{tikzpicture}
\end{subfigure}
\\
\vspace{0.2cm}
\ref{namedmethods4tl}
\caption{Evolution of the aggregate backhaul load and users' satisfaction ratio.}
\label{fig:results-main}
\end{figure*}
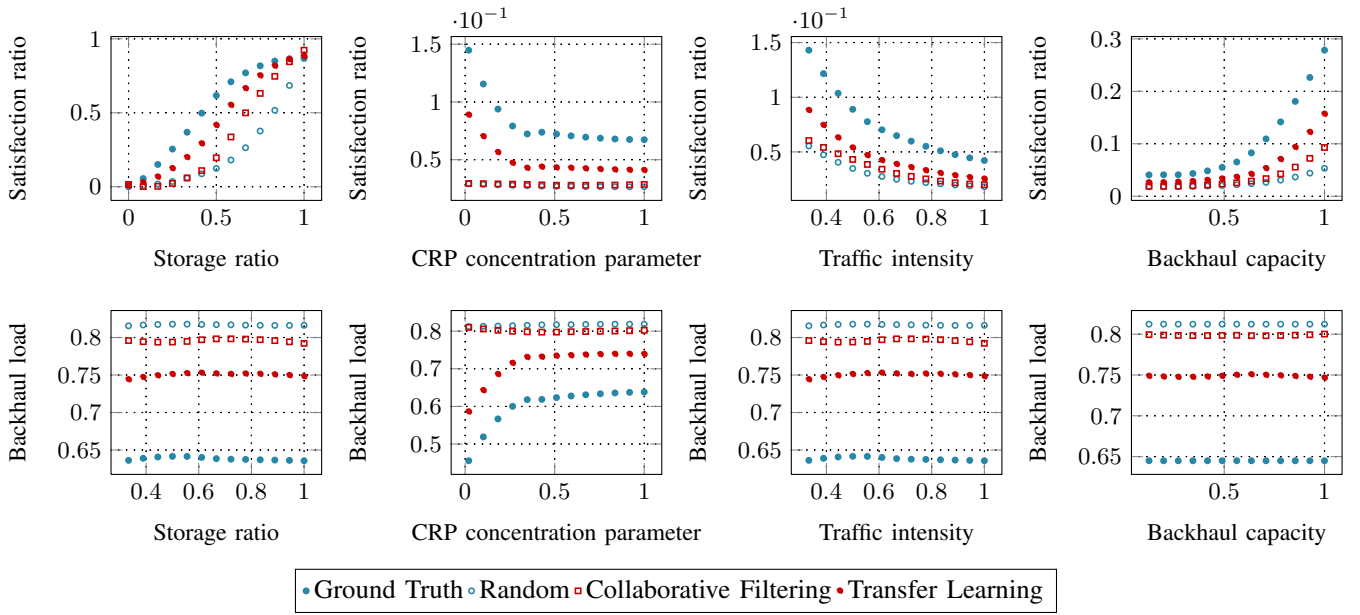

The objective of this section is to validate the effectiveness of the proposed \ac{TL} caching procedure and draw key insights. In particular, we consider the following caching policies for comparison:
\begin{itemize}
	\item[1)]	\emph{Ground Truth}: Given the perfect rating matrix $\bold{P}_{tar}$, the most popular contents are stored greedily.
	\item[2)]	\emph{Random caching} \cite{Bastug2014LivingOnTheEdge}: Contents are cached uniformly at random.
	\item[3)]	\emph{Collaborative Filtering} \cite{Lee2012Cf}: The content popularity matrix $\bold{P}_{tar}$ is estimated via \ac{CF} from a training set with $4\%$ of ratings. Then, the most popular contents are stored accordingly. 
	\item[4)]	\emph{Transfer Learning}: $\bold{P}_{tar}$  and $\bold{P}_{D2D}$ matrices are jointly factorized via \ac{TL} by using a training set with $12\%$ of ratings and perfect user-content correspondence. Then the most popular contents are stored accordingly.
\end{itemize} 
In the numerical setup, having contents cached according to these policies, the \glspl{SBS} serve their users according to a traffic arrival process. This process is drawn from a Poisson process with intensity $\lambda$. The storage size of \glspl{SBS}, content lengths, capacities of non-interfering wireless and backhaul links are assumed to have same constant values individually, in order to showcase the performance of the caching policies. 
The numerical results of users' satisfaction ratio and backhaul load are obtained by averaging out $1000$ Monte-Carlo realizations. The simulation parameters are summarized in Table \ref{tab:simparams}, unless stated otherwise.
\begin{table}[!ht]
\caption{Simulation Parameters}
\label{tab:simparams}
\centering
\scriptsize
\begin{tabular}{c||l||c}
\hline
\textbf{Parameter}	& \textbf{Description}	&	\textbf{Default-Varied Values}  \\
\hline \hline
$M_{tar}$ 	&	Number of \glspl{SBS} 				&	$4$					\\
$N_{tar}$	& 	Number of \glspl{UT}   				& 	$32$ 				\\
$F_{tar}$ 	&	Library size 						& 	$32$ contents 		\\
$L$			&	Content length						& 	$1$ MBit				 \\
$B$			&	Bitrate requirement 					& 	$1$ MBit				 \\
$\sum{C'_m}$&	Total wireless capacity				&	$32$ MBit/s			\\
$T$ 			&	Time slots   						& 	$ 128$ seconds \\
$\alpha$		& 	Zipf parameter 						&	$2$					\\
$\beta$		& 	CRP concentration parameter			&	$2$ - $[2\sim 100]$ 		\\
$\sum{S_m}$		& 	Total storage size 				& 	$6$ - $[0\sim 32]$ MBit	 \\
$\sum{C_m}$		&	Total backhaul capacity  		&	$1$ - $[1\sim 8]$ MBit/s	\\
$\lambda$		& 	Traffic intensity				&	$1$ - $[1\sim 3]$ demand/s \\
\hline
\end{tabular}
\vspace{-0.2cm}
\end{table}

The dynamics of users' satisfaction ratio and backhaul load with respect to the storage size, demand shape in the source domain, traffic intensity and backhaul capacity are given in Fig. \ref{fig:results-main}. The results are normalized to show the various percentage gains, whereas the actual values are shown in Table \ref{tab:simparams}. In the following, we discuss in detail the impact of these parameters.
\subsubsection{Impact of the storage size ($S_m$)}
The storage size is indeed one of the crucial parameter in cache-enabled \glspl{SBS}, and it is expected that higher storage sizes result in better performance in terms of satisfaction ratio and backhaul offloading.  According to this setup, we would like to note that the biggest improvement in satisfaction ratio and decrement in the backhaul load is achieved by the ground truth baseline where the content popularity is perfectly known. The random approach on the other hand has the worst-case performance. The \ac{CF} approach exhibits similar performance as  the random approach due to the cold-start problem, whereas the satisfaction ratio and backhaul offloading gains of \ac{TL}  are   close to the ground truth baseline. In particular, it is shown that the \ac{TL} policy outperforms its \ac{CF} counterpart, with satisfaction and backhaul offloading gains up to $22\%$ and $5\%$ respectively. 
\subsubsection{Impact of the demand shape in the source domain ($\beta$)}
The demand shape in the source domain, characterized by the \ac{CRP} concentration parameter $\beta$  provides meaningful insights to our problem. In fact, as $\beta$ increases, the demand shape tends to be more uniform, requiring higher storage sizes at the \glspl{SBS} to sustain the same performance. In a storage limited case, we see that the satisfaction ratio decreases and the backhaul load increases with the increment of $\beta$. Compared to the \ac{CF} approach, the gains of \ac{TL} are around $6\%$ for the satisfaction gains and $22\%$ for the backhaul offloading. However, the gap between \ac{TL} and \ac{CF} becomes smaller as $\beta$ increases. 
\subsubsection{Impact of the traffic intensity ($\lambda$)}
As the average number of request arrivals per time slot increases, bottlenecks in the network are expected to occur due to the limited resources of \glspl{SBS}, resulting in less satisfaction ratios. This is visible in the high arrival rate regime, whereas the relative backhaul load remains constant. It can be shown that the ground truth caching with perfect knowledge of content popularity outperforms the other policies while the random approach has the worst performance. On the other hand, the performance of \ac{TL} is in between these approaches and has up to $3\%$ satisfaction gains and $18\%$ of backhaul offloading gain compared to the \ac{CF}.
\subsubsection{Impact of the backhaul capacity $(C_m$)}
The total backhaul capacity is  assumed to be sufficiently smaller than the capacity of wireless links. The increment of this capacity clearly results in higher satisfaction ratios in all cases. Note that any content not available in the caches of \glspl{SBS} is delivered via the backhaul. Therefore, increasing the backhaul capacity avoids the bottlenecks during the delivery, thus yielding higher users' satisfaction. On the other hand, the backhaul load remains constant in this setting. It can be seen that \ac{TL} approach has satisfaction ratio gains of up to $6\%$ and backhaul offloading of up to $5\%$ compared to the \ac{CF} approach.

\subsubsection{Impact of source-target correspondence}
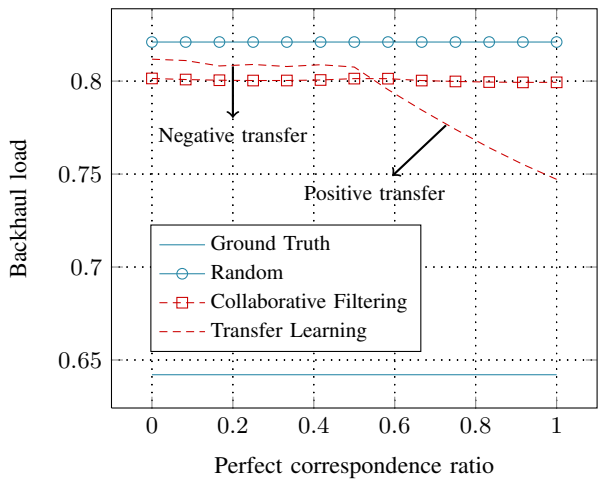
\begin{figure}[ht!]
\begin{subfigure}[t]{0.9\columnwidth}
\begin{tikzpicture}
	\begin{axis}[
		width=\textwidth,
		grid = major,
		cycle list name=laneas4tl,
		legend cell align=left,
		legend style={at={(0.08,0.3)},anchor=west},
		xlabel=Perfect correspondence ratio,
		ylabel=Backhaul load]
		
	\addplot+ plot coordinates {
		(1.6306e-16,	1-0.35793)
		(0.083333,	1-0.35793)
		(0.16667,	1-0.35793)
		(0.25,	1-0.35793)
		(0.33333,	1-0.35793)
		(0.41667,	1-0.35793)
		(0.5,	1-0.35793)
		(0.58333,	1-0.35793)
		(0.66667,	1-0.35793)
		(0.75,	1-0.35793)
		(0.83333,	1-0.35793)
		(0.91667,	1-0.35793)
		(1,	1-0.35793)
	}; 	\addlegendentry{Ground Truth}

	\addplot+ plot coordinates {
		(1.6306e-16,	1-0.17899)
		(0.083333,	1-0.17899)
		(0.16667,	1-0.17899)
		(0.25,	1-0.17899)
		(0.33333,	1-0.17899)
		(0.41667,	1-0.17899)
		(0.5,	1-0.17899)
		(0.58333,	1-0.17899)
		(0.66667,	1-0.17899)
		(0.75,	1-0.17899)
		(0.83333,	1-0.17899)
		(0.91667,	1-0.17899)
		(1,	1-0.17899)
	}; 	\addlegendentry{Random}

	\addplot+ plot coordinates {
		(1.6306e-16,	1-0.19858)
		(0.083333,	1-0.19916)
		(0.16667,	1-0.19953)
		(0.25,	1-0.1997)
		(0.33333,	1-0.19968)
		(0.41667,	1-0.19941)
		(0.5,	1-0.19866)
		(0.58333,	1-0.19869)
		(0.66667,	1-0.19971)
		(0.75,	1-0.20019)
		(0.83333,	1-0.20052)
		(0.91667,	1-0.2007)
		(1,	1-0.20066)
	}; 	\addlegendentry{Collaborative Filtering}

	\addplot+ plot coordinates {
		(1.6306e-16,	1-0.18821)
		(0.083333,	1-0.18895)
		(0.16667,	1-0.19185)
		(0.25,	1-0.19106)
		(0.33333,	1-0.19216)
		(0.41667,	1-0.1912)
		(0.5,	1-0.19245)
		(0.58333,	1-0.20458)
		(0.66667,	1-0.21558)
		(0.75,	1-0.22594)
		(0.83333,	1-0.23582)
		(0.91667,	1-0.24479)
		(1,	1-0.25285)
	}; 	\addlegendentry{Transfer Learning}

	\node[anchor=north] (source) at (axis cs:0.2, 0.818){}; 
	\node (destination) at (axis cs:0.2,  0.77){\footnotesize  Negative transfer}; 
	\draw[thick,->](source)--(destination);
	
	\node[anchor=south] (source) at (axis cs:0.75, 0.776){};
	\node (destination) at (axis cs:0.55,  0.74){\footnotesize  Positive transfer};
	\draw[thick,->](source)--(destination);	
	\end{axis}
\end{tikzpicture}
\end{subfigure}
\caption{Evolution of the backhaul load with respect to the perfect correspondence ratio.}
\label{fig:results-correspondence}
\end{figure}

We have so far assumed that the user/content correspondence between the target and source domains is perfect. This is a strong assumption and such an operation requires a more careful treatment to avoid negative transfer. Here, we relax this assumption by introducing a perfect correspondence ratio. This ratio represents the amount of perfect user/content matching between both source and target domains. A ratio of  $0$ means that $100\%$ of correspondence is done uniformly at random and $1$ is equivalent to the perfect case.
It is shown in Fig. \ref{fig:results-correspondence} that \ac{TL} has a poor performance  in the low values of this ratio, with similar performance as the random caching due to the  negative transfer. However, as this ratio increases, the performance of \ac{TL} improves, outperforming the \ac{CF} with a ratio of $0.58$. This underscores the importance of such an operation for the positive transfer and is left for future work.
\vspace{-0.0cm}
\section{Conclusions}
\label{sec:conclusions}
We proposed a novel transfer learning-based caching procedure which was shown to  yield higher users' satisfaction and backhaul offloading gains overcoming the data sparsity and cold start problems.  Numerical results confirmed that the overall performance can be improved by transferring a judiciously-extracted  knowledge from a source domain to a target domain via \ac{TL}. An interesting future work is assessing the performance of \ac{TL}-based caching using real traces. Another avenue of research is extending the current model to predictive scheduling and predictive offloading.
\bibliographystyle{IEEEtran}
\bibliography{references}
\end{document}